\documentclass[10pt,preprint]{aastex}
\begin{document}

\title{THERMAL DUST EMISSION FROM PROPLYDS, UNRESOLVED DISKS, AND
SHOCKS IN THE ORION NEBULA\altaffilmark{1}}

\author{Nathan Smith\altaffilmark{2,3}, John Bally\altaffilmark{2},
 Ralph Y.\ Shuping\altaffilmark{4}, Mark Morris\altaffilmark{5}, and
 Marc Kassis\altaffilmark{6}
}

\altaffiltext{1}{Based on observations obtained at the Gemini
Observatory, which is operated by the Association of Universities for
Research in Astronomy, Inc., under a cooperative agreement with the
NSF on behalf of the Gemini partnership: the National Science
Foundation (US), the Particle Physics and Astronomy Research Council
(UK), the National Research Council (Canada), CONICYT (Chile), the
Australian Research Council (Australia), CNPq (Brazil), and CONICET
(Argentina).}

\altaffiltext{2}{Center for Astrophysics and Space Astronomy, University of
Colorado, 389 UCB, Boulder, CO 80309}

\altaffiltext{3}{Hubble Fellow; nathans@casa.colorado.edu}

\altaffiltext{4}{University Space Research Association, Stratospheric
  Observatory for IR Astronomy, MS 211-3, Moffett Field, CA 94035}

\altaffiltext{5}{Department of Physics and Astronomy, University of
California, Los Angeles, Los Angeles, CA 90095-1547}

\altaffiltext{6}{Keck Observatory, 65-1120 Mamalahoa Highway, Kamuela,
HI 96743}


\begin{abstract}

We present a new 11.7 $\micron$ mosaic image of the inner Orion nebula
obtained with T-ReCS on Gemini South.  The map covers
2$\farcm$7$\times$1$\farcm$6 with diffraction-limited spatial
resolution of 0$\farcs$35; it includes the BN/KL region, the
Trapezium, and OMC-1 South.  Excluding BN/KL, we detect 91
thermal-infrared point sources, with 27 known proplyds and over 30
``naked'' stars showing no extended structure in {\it Hubble Space
Telescope} ({\it HST}) images.  Within the region we surveyed,
$\sim$80\% of known proplyds show detectable thermal-infrared
emission, almost 40\% of naked stars are detected at 11.7~$\micron$,
and the fraction of all visible sources with 11.7~$\micron$ excess
emission (including both proplyds and stars with unresolved disks) is
roughly 50\%.  These fractions exclude embedded sources.  Thermal dust
emission from stars exhibiting no extended structure in {\it HST}
images is surprising, and means that they have retained circumstellar
dust disks comparable to the size of our solar system. Proplyds and
stars with infrared excess are not distributed randomly in the nebula;
instead, they show a clear anti-correlation in their spatial
distribution, with proplyds clustered close to $\theta^1$C, and other
infrared sources found preferentially farther away. We suspect that
the clustered proplyds trace the youngest $\sim$0.5 Myr age group
associated with the Trapezium, while the more uniformly-distributed
sources trace the older 1--2 Myr population of the Orion Nebula
Cluster. This suggests that small protoplanetary disks persist for a
few Myr in irradiated environments, and hints that hierarchical
sub-clustering has been important on $\sim$30\arcsec\ scales around
the Trapezium.  We detect 11.7~$\micron$ emission from the five
brightest members of the Trapezium ($\theta^1$ABCDE), caused by
free-free stellar wind emission and possibly emission from dusty disks
around companion stars. Within 30$\arcsec$ of $\theta^1$C~Ori, 100\%
of known proplyds are detected at 11.7 $\micron$, and they exhibit
remarkable limb-brightened dust arcs at the collision of the proplyd
mass loss and the wind from $\theta^1$C.  The star $\theta^1$D is
associated with the most prominent mid-IR dust arc of the Ney-Allen
nebula.  We propose that this arc is the consequence of $\theta^1$D
being the closest member of the Trapezium to the background cloud.
Finally, we detect dust emission from Herbig-Haro jets in Orion,
including HH~202, HH~529, HH~513, and HH~514.  This is the first
detection of mid-infrared continuum emission from dust in the body of
a collimated HH jet or bow shock.  The presence of dust implies that
some jet material must be lifted from large radii in the accretion
disk (outside the dust sublimation radius) or entrained from the
circumstellar environment.

\end{abstract}


\section{INTRODUCTION}

Because of its proximity (d$\simeq$460 pc; Bally et al.\ 2000) and
extreme youth, the Orion nebula is a valuable archetype for
investigating low-mass star formation near high-mass stars.
Consequently, Orion is so far the only H~{\sc ii} region where a large
number of externally-illuminated protoplanetary disks can be studied
in detail.  These ``proplyds'' were first distinguished by their
H$\alpha$ and radio emission (Laques \& Vidal 1979; Churchwell et al.\
1987; Garay et al.\ 1987), but their circumstellar disks and ionized
envelopes have been clearly resolved at visual wavelengths in {\it
Hubble Space Telescope} ({\it HST}) images (e.g., O'Dell et al.\
1993a; O'Dell \& Wen 1994; O'Dell \& Wong 1996; O'Dell 1998, 2001;
Chen et al.\ 1998; McCaughrean et al.\ 1998; Bally et al.\ 1998, 2000;
Smith et al.\ 2005).  Meaburn (1988) first pointed out that all the
bright proplyds around the Trapezium contain stellar sources in the
near-infrared.  Over 80\% of known stars in the inner arcminute near
the Trapezium have proplyd features.  Many of them have
spatially-resolved disks seen in silhouette against the background
nebula. Thus, they have retained significant amounts of circumstellar
dust despite their irradiated environment, which motivates a
mid-infrared (IR) census at high spatial resolution.

At IR wavelengths, the OMC-1 region is dominated by hot dust in the
luminous embedded BN/KL nebula.  Not surprisingly, most thermal-IR
imaging studies of Orion have focussed on resolving its dramatic
structure (e.g., Becklin \& Neugebauer 1967; Kleinmann \& Low 1967;
Rieke et al.\ 1973; Downes et al.\ 1981; Wynn-Williams et al.\ 1984;
Gezari 1992; Dougados et al.\ 1993; Gezari et al.\ 1998; Greenhill et
al.\ 2004; Shuping et al.\ 2004).

By comparison, little mid-IR attention has been lavished upon the
Trapezium's vicinity, sometimes called the Ney-Allen nebula in the
mid-IR (Ney \& Allen 1969; Gehrz et al.\ 1975).  The only previous
sub-arcsecond thermal-IR imaging of the Trapezium cluster was by
Hayward et al.\ (1994), who detected several compact sources
associated with proplyds and a series of parabolic arcs surrounding
them, presumably marking the shock between the proplyds' mass-loss and
the stellar wind from $\theta^1$~Ori~C ($\theta^1$C hereafter).
Hayward \& McCaughrean (1997) subsequently detected mid-IR emission
from a few additional proplyds farther from the Trapezium as well (see
also Robberto et al.\ 2002). More recently, Robberto et al.\ (2005)
presented large-scale 10 \& 20 $\micron$ maps of the Orion nebula at
arcsecond resolution, confirming previous extended structures and
providing the first comprehensive list of point sources at mid-IR
wavelengths.  They detected 177 point sources in a
5\arcmin$\times$3$\farcm$5 field, many of these associated with
proplyds.  In the near-IR, $JHKL$ photometry (e.g., Lada et al.\ 2004
and references therein) reveals a large IR-excess disk fraction
(roughly 50\%) among the low-mass stellar population in Orion, as well
as several highly-reddened and deeply-embedded sources that may mark
sites of active ongoing star formation behind the main ionization
front illuminated by the Trapezium.

Our new Gemini images show the same dust structures first noted by
Hayward et al.\ (1994), but with improved spatial resolution and
sensitivity, and we detect numerous additional sources not reported
previously at these wavelengths. These are the first mid-IR images of
a large portion of the Trapezium region obtained with an 8-m class
telescope, and the improved spatial resolution facilitates a
comparison with structures seen in visual-wavelength images obtained
with {\it HST}.  In recent papers (Smith et al.\ 2004b; Smith \& Bally
2005) we reported the discovery of several new IR sources in the OMC-1
South cloud core and a disk around the embedded source IRc9, based on
a subset of the data presented here.  Below we highlight some of the
intricate structure seen in our larger 11.7~$\micron$ mosaic image,
with particular attention to several compact sources (both stars and
proplyds) and the dusty arcuate shocks in the Trapezium region.  The
intricate multiwavelength structure in the BN/KL region will be
discussed in a separate paper.

\section{THERMAL-IR OBSERVATIONS}

Images of the Orion nebula at 11.7 $\micron$
($\Delta\lambda$=11.09-12.22 $\micron$) were obtained on 2004 Jan 25
using T-ReCS on Gemini South.  T-ReCS is the facility mid-IR imager
and spectrograph with a 320$\times$240 pixel Si:As IBC array, a pixel
scale on the 8m Gemini South telescope of 0$\farcs$089, and a
resulting field-of-view of 28$\farcs$5$\times$21$\farcs$4.  The
observations were taken with a 15$\arcsec$ east-west chop throw.
Since the throw is much smaller than the extent of the nebula, we took
a series of adjacent images, starting on relatively blank sky to the
west, and stepping by 15$\arcsec$ per pointing at a constant
declination.  To define the reference sky to subtract from each
position, we took the minimum of two frames on either side of the
position of interest.  This allowed effective removal of point sources
in the reference sky frames, but some nebular emission persisted where
both adjacent pointings contained diffuse emission at the same
position on the array, or where a point source and bright diffuse
emission overlapped in the two reference frames adjacent to the array
position being considered.  In those cases, we set the residual
emission in the sky frame by interpolating over the point source. This
technique worked sufficiently well to reveal structures smaller than
15$\arcsec$ in size, but degraded the image quality somewhat because
of small pointing errors in the off-source beam (the telescope cannot
guide while observing the reference sky position).  Individual
sky-subtracted frames were then combined to make a larger mosaic
image, using point sources in each frame as tie points for spatial
alignment.

Figure 1 shows the resulting 11.7~$\micron$ mosaic image.  Table 1
lists J2000 coordinates for IR point sources, as well as point-source
photometry at 11.7 $\micron$ measured in a 0$\farcs$9 radius synthetic
aperture.  These coordinates were measured relative to $\theta^1$C
assuming its position is $\alpha_{2000}$=5$^{\rm h}$35$^{\rm
m}$16$\fs$46, $\delta_{2000}$=$-$5$\arcdeg$23$\arcmin$23$\farcs$0
(McCaughrean \& Stauffer 1994).  The relative positional uncertainty
is better than 1 pixel ($\sim$0$\farcs$09) near the center of the
field but increases to about 5 pixels (0$\farcs$45) near the upper and
lower edges.  Flux densities in Table 1 were measured with respect to
the secondary standard star HD~32887, adopting the values tabulated by
Cohen et al.\ (1999).  Point-source sensitivity was approximately 10.5
mJy (1$\sigma$) at 11.7 $\micron$.  The measured FWHM of point sources
in our images was roughly 0$\farcs$35, consistent with the expected
diffraction limit at this wavelength. Flux uncertainty for the
brighter sources is dominated by $\pm$5--10\% uncertainty in the
calibration stars.  However, because we were limited by a relatively
small chopper throw of only 15$\arcsec$, at some positions where
bright and complex diffuse emission fell in the reference beam
(i.e. near the Trapezium) the photometric accuracy may be as poor as
$\pm$30\%, and a few sources listed in parenthesis had additional
uncertainty due to a close neighbor or uneven background.

In \S 3.3 we supplement our monochromatic Gemini/T-ReCS map with data
obtained with the MIRSI instrument on the NASA Infrared Telescope
Facility (IRTF).  We used the flux ratio in 11.7 and 24.5~$\micron$
images to deduce the grain color temperature.  Details of the data
reduction for the MIRSI images can be found in Kassis et al.\ (2005).

\section{RESULTS}

Our 11.7 $\micron$ map has the best combination of high spatial
resolution, sensitivity, and areal coverage that has yet been used to
image Orion in the mid-IR.  This map has a smaller field of view than
a recent mid-IR survey by Robberto et al.\ (2005), but the higher
spatial resolution and sensitivity of our study makes the two datasets
complementary, especially near the crowded Trapezium where spatial
resolution is key.  Our higher spatial resolution also facilitates a
detailed comparison between dust emission structures and emission-line
features seen in narrowband optical {\it HST} images.  In our T-ReCS
survey, many new point sources and diffuse structures are detected,
some close pairs of sources are resolved for the first time, and
previously-identified nebular features are revealed in greater detail,
as discussed below.

\subsection{Proplyds and Stars}

Within the boundaries of Figure 1, we detect 91 point sources (see
Table 1; this excludes sources assumed to be part of the BN/KL complex
and faint extended proplyds like 141-301).  Labels in Figures 1 and 2
identify 11.7~$\micron$ sources which are known to be proplyds because
of extended structure seen in {\it HST} images, while IR sources not
associated with proplyds are circled in Figure 2.  Some of these
non-proplyd IR sources have no visual I.D. and are likely to be
embedded sources.  Table 2 summarizes source counts and IR excess
fractions for all sources with visual-wavelength counterparts (both
proplyds and ``naked'' stars showing no extended structure in {\it
HST} images, taken from the list compiled by O'Dell \& Wong 1996).
The fraction of visible sources showing excess 11.7~$\micron$ emission
is roughly 50\%, which agrees with the IR-excess disk fraction
measured for the low-mass stellar population in the nebula by Lada et
al.\ (2004).

We detected nearly all the known proplyds (O'Dell \& Wong 1996; Bally
et al.\ 2000) in our survey region, except for 161-328, HST 1, 10, 12,
13, 16, and 17, so that about 80\% of known proplyds have detectable
11.7~$\micron$ emission.  The non-detections of HST~1, 10, 16, and 17
need to be qualified, however, as they fell in the 15$\arcsec$-wide
portion at the left edge of Figure 1 which was only observed in the
reference sky position and is of considerably lower quality (note that
faint extended emission from HST~1 and 10 was indeed detected in
N-band images by Robberto et al.\ 2002).

Perhaps even more interesting than the detection of proplyds is the
11.7 $\micron$ emission from several ``naked'' stars with no extended
proplyd structure in {\it HST} images, some of which are brighter than
nearby proplyds at 11.7 $\micron$.  These comprise about half the IR
point sources circled in Figure 2.  The mid-IR emission from most of
the weaker sources is detected here for the first time. We detect 27
new 11.7 $\micron$ sources not found previously in $N$-band images by
Robberto et al.\ 2005 (new embedded sources in OMC-1 South were
discussed in a previous paper; Smith et al.\ 2004b).  In the last
column of Table 1 we list numbers assigned to various single point
sources detected previously by Robberto et al.\ (2005).  Except in the
case of the luminous Trapezium stars themselves, the detectable excess
mid-IR emission implies that these stars have retained dust disks,
despite the lack of material in observable proplyd envelopes.  Since
no extended structure is seen by {\it HST} (either as ionized
envelopes or as silhouette disks), these must be very compact remnant
dust disks -- roughly the size of Pluto's orbit or smaller. Since they
do not show extended structure in {\it HST} images, the disks are
either depleted of gas so that they do not have significant
photoevaporating envelopes, or their evaporation rates are so low that
their ionized envelopes remain unseen in optical emission lines by
{\it HST}.  There is no apparent trend of 11.7 $\micron$ flux with
separation from $\theta^1$C.  The apparent thermal-IR flux of an
externally-irradiated disk will also depend on its size, its tilt
angle with respect to the ionizing source, and its inclination from
our line of sight (see Robberto et al.\ 2002).  Therefore, the dust
may be heated internally by the central star in each system.  Many of
these stars also show excess L' band emission (Lada et al.\ 2004), and
a detailed study of their spectral energy distributions (SEDs),
including additional multi-filter mid-IR emission, would be worthwhile
to further constrain the properties of the disks.  Other implications
of our detection of these mid-IR point sources are discussed further
in \S 4.1.

All of the main Trapezium stars ($\theta^1$A, B, C, D, and E) are
detected at 11.7 $\micron$, but they are faint (Figure 3).  Their
mid-IR fluxes are roughly consistent with expected photospheric
emission, perhaps enhanced by moderate free-free emission from their
stellar winds. However, the high multiplicity fraction in the
Trapezium\footnote{The five Trapezium stars $\theta^1$ABCDE actually
contain at least 13 individual stars. $\theta^1$A and $\theta^1$C each
have at least one spatially-resolved companion, $\theta^1$B is
composed of at least four resolved stars, and both $\theta^1$A and
$\theta^1$B are eclipsing binaries (Weigelt et al.\ 1999; Simon et
al.\ 1999; Schertl et al.\ 2003).} raises the possibility that the IR
emission may originate in dusty disks around companion stars.  Indeed,
Schertl et al.\ (2003) found near-IR excess emission associated with
companions of $\theta^1$B and $\theta^1$C, and relatively high
extinction toward companions of $\theta^1$A and $\theta^1$B.  External
irradiation of such disks less than $\sim$500 AU from a luminous star
should result in very hot dust around these sources, which should be
distinguishable from free-free or photospheric emission in the SEDs of
$\theta^1$ABCDE.  $\theta^1$B has a known proplyd companion (161-307;
Bally et al.\ 2000) that is clearly spatially resolved for the first
time at thermal-IR wavelengths in Figure 3.  Of the main Trapezium
stars, only $\theta^1$D has no known companions, and the remarkable
peculiarities of its environment are discussed below in \S 3.3.

Not all proplyds detected at 11.7~$\micron$ are point sources in the
IR; some show complex structure or extended diffuse emission.  Figure
4 shows two interesting examples: the large proplyd 141-301, and the
proplyd 165-235 with the associated microjet HH~513.  In 141-301, the
thermal IR emission comes not from a disk concentrated around the
central star, but from the long dark tail trailing behind the proplyd
in the direction away from $\theta^1$C.  This proplyd is unusual
compared to many other of Orion's bright proplyds and silhouette disks
in that the proplyd is very large and the full extent of the tail is
seen as a silhouette object.  This large size and full-body silhouette
are reminiscent of some of the large proplyd candidates seen in the
Carina nebula (Smith et al.\ 2003).  Whether the tadpole shapes of
these objects and 141-301 are caused by true evaporating
protoplanetary disks or instead by evaporating cometary globules is
not yet clear.  The T-ReCS images also spatially resolve close pairs
of sources containing proplyds, like $\theta^1$B and its companion
161-307, and the proplyd HST3 plus its nearby companion.  Although
they are not detected in our images because they reside in the
poor-quality region at the east edge of our map, HST 1 and HST 10 also
show extended mid-IR emission (Robberto et al.\ 2002).

\subsection{Dust Arcs}

Figure 3$a$ shows the remarkable thin limb-brightened arcs around the
proplyds LV 1, 2, 3, 4, and 5, 157-323, 158-326, and 166-316.  Their
mid-IR emission was noted earlier by Hayward et al.\ (1994), but their
structure is revealed here in greater detail.  Many of these arcs are
also seen in H$\alpha$ and [O~{\sc iii}] $\lambda$5007 emission
(Figure 3$b$; see also Bally et al.\ 2000), implying that they mark
the collision between the stellar wind of $\theta^1$C and the ionized
proplyd outflow (e.g., Henney \& Arthur 1997).  In all cases, the
parabolic arcs point in the general direction of $\theta^1$C.

The source just west of $\theta^1$C (163-323) shows an unusual
envelope with a radius of 1$\farcs$5; it has no known evidence for an
ionized proplyd cusp in {\it HST} images, but such emission would be
difficult to detect as this source is embedded in the glare from
$\theta^1$C.  Unlike the other mid-IR arcs, this star's envelope
appears roughly spherical, despite its proximity to $\theta^1$C.
Robberto et al.\ (2002) interpret this source (also named SC3; Hayward
et al.\ 1994) as a proplyd surrounding a face-on disk located behind
$\theta^1$C and projected near it on the sky.  An alternative
interpretation might be that it is close to or embedded within the
photodissociation region (PDR) behind $\theta^1$C, much as we suggest
below for the arc around $\theta^1$D.  However, the fact that the
central star is a bright IR source suggests that it does indeed harbor
a warm circumstellar disk.

The dust arcs associated with LV1 and $\theta^1$D are larger and much
brighter than the other proplyd arcs in Figure 3$a$.  Their
corresponding 11.7~$\micron$/H$\alpha$ flux ratios are much higher as
well; the arc around LV1 is only faintly visible in {\it HST} images,
and the large arc around $\theta^1$D is not seen at all in H$\alpha$
or [O~{\sc iii}] (Fig.\ 3$b$).  In the case of $\theta^1$D, at least,
we propose that the dust arc emitting at 11.7~$\micron$ has a
different origin than the arcs around the other Trapezium proplyds, as
discussed below.

The dust arc around LV1 is unusual in that it is much brighter than
any of the other proplyd arcs.  LV1 is also by far the brightest
proplyd at 11.7~$\micron$.  These clues might suggest that it is
closer to $\theta^1$C than the other proplyds, despite its greater
separation than some others (like LV4 and 163-323) as seen projected
on the sky.  This view is supported by the much warmer dust
temperatures around LV1 of roughly 160--170 K (Figure 5), as compared
to $\sim$130 K for other proplyds with a similar apparent separation
from $\theta^1$C.  However, this view would not easily explain the
much fainter H$\alpha$ and [O~{\sc iii}] emission from the large arc
around LV1, since the ionizing UV radiation should be more intense as
well.  Another interpretation might be that the outflow from LV1 has a
higher dust content than the other proplyd flows, or that it contains
smaller and hotter grains, making it appear brighter at
11.7~$\micron$.  These questions can potentially be answered with more
detailed multi-filter mid-IR imaging or mid-IR spectroscopy.  Yet, it
is worth pointing out that while the LV1 arc is somewhat different
from the other proplyd arcs, its dust cannot have the same PDR origin
that we attribute to the Ney-Allen dust in the following section; the
LV1 arc does indeed point directly toward $\theta^1$C while the
Ney-Allen arc does not, and LV1 does not contain a comparable luminous
heat source.

\subsection{$\theta^1$D and the Ney-Allen Nebula}

The environment around $\theta^1$D is particularly interesting.
Within a radius of about 0$\farcs$6$-$0$\farcs$8, it shows a 0.1$-$0.2
Jy arcsec$^{-2}$ deficit of thermal-IR emission compared to its
surroundings, perhaps indicating a dust-free cavity around the star.
This is visible in Figures 1, 2, and 3$a$, but it is shown more
clearly by the tracing through the star in Figure 6.  This empty ring
around the star is not an image processing artifact, since we have
applied no image enhancement like unsharp masking or maximum-entropy
deconvolution to our T-ReCS image, which can sometimes manufacture
similar rings.

Does this indicate a real dust cavity around $\theta^1$D?  Any
luminous star will have a zone around it where dust would be too hot
to remain as solid grains; $\theta^1$D is a normal B0.5 V star with
L=5$\times$10$^4$~L$_{\odot}$.  For normal dust grains that condense
at temperatures of roughly 1000 K, assuming a typical grain radius of
$a\simeq$0.1~$\micron$ (so that $Q_{\rm em}/Q_{\rm abs}\simeq$100),
dust should be able to survive no closer than 180 AU or 0$\farcs$4
from $\theta^1$D.  Since this is in rough agreement with the size of
the ``cavity'' we observe, one could surmize that either 1)
$\theta^1$D produces dust in its own stellar wind beyond about 200 AU
from the star, or 2) $\theta^1$D is embedded within a dusty
environment and has destroyed that dust in its immediate vicinity.
Since main-sequence O and early B stars do not normally have dusty
stellar winds, and because efficient dust formation requires
relatively high densities unlikely to be found in a stellar wind at a
distance as large as 180 AU, the second option seems more likely.
Locating $\theta^1$D within a dusty environment has important bearing
on the discussion below.

$\theta^1$D is close to the center-of-curvature of a large
2--6$\arcsec$-radius arc that forms the most prominent feature of the
Ney-Allen nebula (see Fig.\ 3$a$).  Robberto et al. (2005) proposed
the existence of a massive evaporating circumstellar disk around
$\theta^1$D to explain the formation of this arc.  They noted a
possible connection between this IR-excess disk around $\theta^1$D and
the fact that $\theta^1$D is the only massive Trapezium member without
a known binary or multiple companion star, suggesting that these were
mutually exclusive.  Our higher resolution mid-IR images, however,
show no such dust disk around $\theta^1$D larger than $\sim$50 AU --
instead they show what appears to be a cavity around the star, with
the weak IR emission from $\theta^1$D itself being consistent with
photospheric emission only.  Following the discussion above, it is
unlikely that $\theta^1$D harbors any dust disk at all, since an
unresolved disk responsible for the observed 11.7~$\micron$ emission
would be well inside the dust sublimation radius.  Thus, the origin of
the Ney-Allen arc must have some other explanation.

This arc cannot be discerned in visual-wavelength images, and its
origin is mysterious.  There is no source other than $\theta^1$D that
it could plausibly be associated with.  $\theta^1$D is offset from its
center-of-curvature by about half of the arc's radius, and unlike the
other arcs in the Trapezium, its apex does not point exactly toward
$\theta^1$C.  However, it is intriguing that the apex of the arc does
point in roughly the same direction as the proper motion of
$\theta^1$D itself, measured as 0.5 mas yr$^{-1}$ (a tangential
velocity of $\sim$1 km s$^{-1}$) at P.A.$\simeq$217\arcdeg\ (van
Altena et al.\ 1988).  This direction is shown by the arrow near
$\theta^1$D in Fig.\ 3$a$.  Given the very bright thermal-IR emission
and this structure's invisibility at optical wavelengths, we speculate
that the Ney-Allen arc may arise as the stellar wind of $\theta^1$D
sweeps up dust in the main PDR behind the Orion nebula, and that dust
in the Ney-Allen arc is heated by $\theta^1$D itself, not $\theta^1$C
like all the other Trapezium dust arcs.

This would require that $\theta^1$D be closer to the background
molecular cloud than the other Trapezium stars, which is plausible for
two reasons:

1. $\theta^1$D shows an absorption component at 21 km s$^{-1}$ that is
not seen in optical spectra of the other Trapezium stars (O'Dell et
al.\ 1993b), implying that it is the closest to the ionization front
(O'Dell 2001).

2.  The very bright 11.7 $\micron$ emission from a large area
associated with the Ney-Allen nebula has an elevated color temperature
compared to surrounding regions in the Trapezium.  Figure 5 shows a
color temperature map deduced from the 11.7/24.5~$\micron$ flux ratio
in images obtained with the MIRSI camera at the IRTF (Kassis et al.\
2005).  This map was made assuming a dust emissivity $Q_{\rm
  em}\propto\lambda^{-1}$, revealing grain temperatures of roughly 155
K at the position of the large Ney-Allen arc.  Since the local dust
temperature peaks near $\theta^1$D instead of near the much more
luminous star $\theta^1$C, this requires that $\theta^1$D is located
much closer to the dust than $\theta^1$C is.  The fact that
20~$\micron$ emission is more spatially extended in the direction away
from $\theta^1$D than emission at shorter wavelengths is (Robberto et
al.\ 2005) also suggests that it is the heat source for the arc,
rather than $\theta^1$C.  For a grain temperature of 155 K and the
same assumptions about grain properties as above, this dust should be
found at a distance of $\sim$7500 AU from $\theta^1$D.  This value is
much smaller than the expected distance between the ionizing star
$\theta^1$C and the ionization front behind it, which is roughly
0.2--0.25 pc or 40,000--50,000 AU (Wen \& O'Dell 1995).

\subsection{HH Jets}

Our T-ReCS/Gemini mosaic of the Orion nebula provides the first
unambiguous detection, to our knowledge, of mid-IR continuum emission
from dust in the bow shocks and jets of Herbig-Haro (HH) objects.
This is significant, since dust is frequently assumed to be destroyed
in such shocks (e.g., Reipurth \& Bally 2001).  While mid-IR line
emission from H$_2$ was detected by the {\it Infrared Spce
Observatory} ({\it ISO}) satellite in HH~54 (Neufeld et al.\ 1998) and
HH~337 (Noriega-Crespo et al.\ 1998), mid-IR thermal continuum
emission from hot dust was not.  Molinari et al.\ (1999) detected
far-IR continuum emission from cool dust associated with HH~7, but
those observations had low spatial resolution compared to ground-based
images, so it is not clear if the dust is located in the outflow or
around it.

We have clearly detected 11.7~$\micron$ emission coincident with
bright ionized gas in HH~202, 513, 514, and 529 (Figs.\ 3, 4, and 7).
The 11.7 $\micron$ emission is most likely thermal dust emission,
since the 11.7 $\micron$ filter does not transmit [Ne~{\sc ii}] 12.8
$\micron$ or [S~{\sc iv}] 10.4 $\micron$, and hydrocarbon emission
would not be expected in an irradiated jet or shock in the interior of
an H~{\sc ii} region.  However, spatially-resolved mid-IR spectra
would obviously help to check this assertion.

HH~529 and HH~202 are both associated with prominent jets emerging
from OMC-1~S (Smith et al.\ 2004b; Doi et al.\ 2004; O'Dell \& Doi
2003; Bally et al.\ 2000), and both show detectable 11.7~$\micron$
emission in Figure 7.  We also detect dust emission in the jets from
two proplyds: the highly collimated HH~514 jet from the proplyd HST2
(Fig.\ 3) and the HH~513 microjet from the proplyd 165-235 (Fig.\ 4$c$
and $d$).

In every case where we detect dust in HH jets, we only detect
11.7~$\micron$ emission on the side of the jet facing $\theta^1$C,
while we do not detect the counter-jets in any of these HH objects.
This suggests that the dust is either heated directly by $\theta^1$C,
or is heated indirectly and locally by trapped Ly$\alpha$ from the
dense ionized gas in each object.  We find a direct spatial
correlation between the locations of ionized gas traced by either
H$\alpha$ or [O~{\sc iii}] and the dust seen in mid-IR emission.  The
most impressive example is HH~529 in Figure 7$a$ and $b$, where the
dust emission is seen both behind the leading bow shock and in
internal shocks or density enhancements within the body of the jet.
In the brightest part of HH~529 behind the leading bow shock, we
measure a background-subtracted specific intensity at 11.7~$\micron$
of $I_{\nu}\simeq$0.61 Jy arcsec$^{-2}$.  Dust emission is also seen
in the collimated body of the HH~514 jet from HST2.  Finding mid-IR
emission from hot dust in the body of a collimated jet far from a main
shock front is an important clue to the origin of the dust in HH
flows.  Specifically, it implies that the dust was entrained in the
flow early-on, instead of having formed in the dense cooling zone
behind a reverse shock or bow shock.  This is discussed in more detail
in \S 4.2.

\section{DISCUSSION}

\subsection{Spatial Segregation and Subclustering, and the Lifetime of
Remnant Dust Disks in Irradiated Environments}

Our 11.7 $\micron$ survey of the inner parts of the Orion nebula with
T-ReCS on Gemini South has revealed a number of mid-IR point sources,
which come in three flavors: 1) known proplyds, 2) visible stars with
no extended structure in {\it HST} images (``naked'' stars), and 3)
embedded IR sources with no visual identification.  Of these three
types, the mid-IR emission from ``naked'' stars is perhaps the most
surprising and interesting from the point of view of understanding the
history of recent star formation in this region.  We expected to see
mid-IR emission from proplyds -- which obviously are still surrounded
by gas and dust envelopes, many showing clear silhouette disks or
microjets (e.g., Bally et al.\ 2000) -- and we expect mid-IR emission
from sources embedded within the background molecular cloud.

The mid-IR emission from naked stars, on the other hand, is more
unexpected, indicating the presence of remnant dust disks comparable
to the size of our solar system (optically thick dust disks would need
to be smaller than about 0$\farcs$1 or $\sim$50 AU to escape detection
by {\it HST}).  One might surmise {\it a priori} that these objects
are older than the proplyds, since their smaller disks seem to be in a
more advanced stage of evaporation (the source of that
photoevaporation may have been Balmer continuum radiation from other
luminous stars in the Orion Nebula Cluster, such as $\theta^2$A and
$\theta^2$B).  If true, the age of their central stars may give
important clues to the lifetimes of disks in irradiated environments,
and their spatial distribution within the Orion Nebula Cluster (ONC)
might hold important clues to the geometry of its recent star
formation history.  Hillenbrand (1997) found that stars in the ONC
have a range of ages from 0.5--2 Myr, with the younger 0.5 Myr
population concentrated toward the center, within $\sim$0.3 pc of
$\theta^1$C, and the older 1--2 Myr population distributed more
uniformly over larger radii.  Thus, if the unresolved disks really
have been exposed to evaporating UV radiation for a longer time than
the proplyds, we might expect the proplyds and unresolved disks to
follow a spatial trend akin to the age segregation of the ONC.

Indeed, we find that the three types of mid-IR point sources are not
distributed randomly in the Orion nebula.  Figure 8 shows the
locations of proplyds (filled circles), naked stars with mid-IR
emission (unfilled circles), and embedded mid-IR sources (X's).
Figure 8 gives a clear impression that proplyds are clustered
preferentially near $\theta^1$C, while the remaining mid-IR sources
seem to have a smoother spatial distribution.  This is {\it not}
simply an effect of otherwise similar disks being more severely
evaporated near $\theta^1$C, because a few bright proplyds are seen at
larger distances than naked stars, and the naked stars do not have
spatially-resolved silhouette disks like many proplyds.  These really
are two different populations of objects, reinforcing the idea that
the two types of objects may have different ages.

Figure 9 shows a more quantitative description of the spatial
distribution of these three types of sources, where we have plotted
the number of each type of source as a function of projected
separation from $\theta^1$C (we do not include the massive members
$\theta^1$ABCDE in Fig.\ 9).  {\it There is a clear anti-correlation
between the spatial distribution of proplyds and other types of mid-IR
point sources.}  There appears to be a boundary at a radius of roughly
20\arcsec\ from $\theta^1$C, inside of which almost all mid-IR sources
are proplyds, and outside of which the number of proplyds drops while
the number of non-proplyd point sources rises sharply.  The
populations of naked stars and embedded sources are indistinguishable
in Figure 9, and their sum (naked stars plus embedded sources; the
dashed histogram in Figure 9) shows the anticorrelation even more
vividly.  We did not attempt to fit the trend of the number of each
type of source as a function of radius, since our survey is
incomplete, and only has full coverage over 2$\pi$ radians in azimuth
out to $\sim$20\arcsec\ (0.05 pc).  A larger and more complete survey
that extends farther to the east and southeast might find additional
proplyds or other point sources beyond 20\arcsec\ from $\theta^1$C,
but the dearth of naked stars with mid-IR emission within 20\arcsec\
is real, and somewhat perplexing.

In general, we find that the proplyds detected in the mid-IR are
highly concentrated within 0.05 pc of $\theta^1$C, while other mid-IR
point sources are excluded from this inner region and are distributed
more uniformly outside of it.  This distribution echoes the
segregation by age in the ONC, with younger (0.5 Myr old) stars
heavily concentrated in the center of the cluster, and older (1--2
Myr) stars found over a larger area (Hillenbrand 1997).  This provides
circumstantial evidence that the proplyds near $\theta^1$C are, in
fact, among the youngest members of the Trapezium cluster, while the
mid-IR point sources associated with naked stars and embedded sources
may be part of the older ONC population.

This implies that protoplanetary disks may be long-lived.  Proplyd
lifetimes are often estimated to be only 10$^4$--10$^5$ yr (e.g.,
O'Dell 2001; Bally et al.\ 1998; Henney \& O'Dell 1999), but these
estimates extrapolate from their {\it current} mass-loss rates and
{\it current} disk masses.  As a protoplanetary disk continues to
evaporate and shrinks in size, it's geometric cross section for
absorbing UV radiation also shrinks.  As the disk's outer radius
decreases to the appropriate gravitational radius

\begin{displaymath}
R_G \ \simeq \ \frac{GM}{c_i^2} \, ,
\end{displaymath}

\noindent where $c_I \ \simeq$ 3 km s$^{-1}$ is the sound speed in the
PDR (heated by soft UV radiation) or $c_{II} \ \simeq$ 11 km s$^{-1}$
in the photoionized gas, further evaporation becomes limited by
viscous transport from $R < R_G$ to larger radii where it can escape.
Therefore, as disks shrink to $R\la$40 AU (for 1 M$_{\odot}$), their
evaporative evolution becomes dependent on disk viscosity (Matsuyama,
Johnstone, \& Hartmann 2003).  Thus, in advanced stages of disk
evolution, the corresponding evaporation mass-loss rate declines,
potentially prolonging the disk lifetime.

Throop \& Bally (2005) have recently proposed that as evaporation
continues, the protoplanetary disks will preferentially shed their gas
and smallest grains, leaving behind larger grains that settle to the
disk midplane, thereby aiding in planetesimal (and planet) formation.
We suspect that the mid-IR emitting disks surrounding ``naked'' stars
in Orion are a manifestation of these more advanced phases of
protoplanetary disk evaporation in H~{\sc ii} regions.  Their
association with the older and more extended population of the ONC
might then suggest that the inner portions (5--20 AU) of
protoplanetary disks are resiliant, surviving for at least 1--2 Myr.
If this conjecture is true, then detailed study of these sources holds
enormous potential for learning about a critical phase of planet
formation.  Constraining their disk masses and grain properties could
provide a direct test of the scenario proposed by Throop \& Bally
(2005).  An alternative view could be that some of the ``naked'' stars
were ejected from the core that formed the Trapezium, similar to the
process at work in the simulations of Bate \& Bonnell (2005).  In that
case, their surprisingly small disks for their young age might be the
result of tidal stripping in the ejection process (e.g., Reipurth
2000).  The spatial distribution of these ``naked'' stars over a
larger area of the nebula might shed light on this question.

Additionally, the concentration of proplyds near $\theta^1$C suggests
that hierarchical subclustering has been very important in the
formation of the larger ONC, if the proplyds and the Trapezium stars
really are the youngest objects seen at visual wavelengths in Orion.
Hillenbrand \& Hartmann (1998) pointed out that the ONC is elongated
in the north-south direction, like the molecular cloud behind it
(Johnstone \& Bally 1999), implying that the ONC is young enough that
it is not yet dynamically relaxed and still bears an imprint of the
geometry of its birth environment.  Thus, it seems plausible that the
proplyds and Trapezium stars all formed recently from the same
localized dense cloud core, distinct from the rest of the ONC cluster
--- much as the BN/KL and OMC-1 South events will produce subclusters
younger than the Trapezium when they have cleared away their natal
material.  This is an important clue to the enduring puzzle of why
proplyds are so common in the core of the Orion nebula, but are not
seen elsewhere in the sky -- i.e., the Trapezium and the surrounding
proplyds may be a uniquely young remnant of the original cloud core
that spawned them, where the UV radiation was just turned-on only
0.5--1 Myr ago.  Proplyds may not be seen in other nearby H~{\sc ii}
regions with massive O-type stars because they are not young enough,
and their proplyds have already been mostly evaporated.  More distant
proplyds seen outside the immediate vicinity of $\theta^1$C may have
recently emerged from their own cloud core, or may have been ejected
from the core that formed the Trapezium (e.g., Reipurth 2000; Bate \&
Bonnell 2005).

Finally, Figure 10 shows that even the spatial distribution of the
remaining non-proplyd sources seen in the mid-IR is not entirely
random.  Specifically, embedded sources (X's in Fig.\ 10) seem to be
found preferentially near or on the Trapezium-facing side of the
densest parts of the OMC-1 cloud core, seen in sub-mm emission from
cool dust associated with the ``integral shaped filament'' that runs
through Orion (Johnstone \& Bally 1999).  Lada et al.\ (2004) found a
similar result for the spatial distribution of embedded sources with
excess L-band emission.  It is unclear whether these embedded sources
are still in the early phases of formation within their natal
material, or are instead members of the older 1--2 Myr population that
happen to be engulfed by the dense cloud core that bisects the larger
ONC.  A more thorough mid-IR survey over a larger area, perhaps with
{\it Spitzer}, could help clarify the nature of these embedded
sources.

\subsection{Mid-IR Properties of Point Sources}

Unfortunately, our T-ReCS survey of Orion is monochromatic over most
of the field, limiting its potential for investigating dust properties
(temperature, optical depth, grain size) in the sources we detected.
The nearest wavelength at which a complete survey has been done with
comparable sensitivity and resolution is the L-band study of Lada et
al.\ (2004) using the {\it VLT}.  Interpreting the comparison between
3 and 12 $\micron$ is hampered by the partial contribution of
photospheric emission in the L-band, and the fact that several of the
embedded sources we detected at 11.7 $\micron$ were not detected in
the L-band.  Nevertheless, a preliminary comparison of these two
wavelengths does give some limited insight.

Figure 11 shows a pseudo color-magnitude diagram for the sources in
Table 1, comparing the 11.7 $\micron$ flux to the 11.7
$\micron$/L-band flux ratio.  In Figure 11, the different source
categories are plotted with the same symbols as in Figures 8 and 10.

A handful of proplyds and embedded sources have very strong mid-IR
excess emission (they are found at the upper right in Fig.\ 11).  The
reddest sources are LV1, LV4, and 163-323, all of which are very close
to $\theta^1$C.  Aside from these, most point sources detected in our
survey have similar 3-12 $\micron$ colors, regardless of the type of
source.  With a flux ratio near 1, most of these sources would be
considered Class I, but a more complete investigation of their SEDs
over a wider wavelength range is necessary.  The similar loci for most
of the sources in Figure 11, regardless of the type of source,
suggests no severe differences in grain sizes and temperatures.  While
this does not confirm the age segregation proposed in the previous
section, it is not necessarily inconsistent with it; longer wavelength
data may be needed to detect the largest and coolest dust grains,
which may be preferentially present in the older systems.

In viewing Figure 11, it is useful to recall that it only traces about
half the stellar sources known in the area of the Orion nebula that we
covered -- the remaining stars not detected at 11.7 $\micron$ would
presumably reside at the lower left corner of this figure with weak
11.7 $\micron$ excess and L-band fluxes dominated by photospheric
emission.

\subsection{HH Jets and the Trapezium Dust Arcs -- Where Does the Dust
Come From?}

As noted in \S 3.4, evidence for dust in HH objects is scarce
(Reipurth \& Bally 2001; Henney et al. 1994). Yet, we have detected
11.7 $\micron$ continuum emission from warm dust in several irradiated
HH jets and in arcs around proplyds near $\theta^1$C.  In both cases,
the dust is near enough to $\theta^1$C that it can be heated directly
by UV radiation or indirectly by re-radiation from the ionized gas.
This suggests that dust could be common in other HH jets as well, but
in quiescent environments without a strong external UV source, the
dust is not heated sufficiently to emit strongly in the mid-IR.  While
we see this dust emission in the Orion nebula, the origin of the dust
in these outflows is not obvious, with four potential sources to
consider:

1) In the HH jets, the dust may be entrained at the origin of the jet,
   implying that the jet is lauched from the accretion disk at a
   radius larger than the dust sublimation radius. In the Trapezium
   dust arcs, the dust may be entrained in the photoevaporative flow
   off the surface of the central proplyd disk.

2) Dust may be entrained from the ambient material surrounding the jet
   or proplyd.  This process seems to be at work in HH~444 near
   $\sigma$ Ori, as demonstrated by Andrews et al.\ (2004).

3) The dust may form in the dense post-shock cooling zone, either from
   jet or proplyd material after passing through the reverse shock, or
   from ambient material that has passed through the forward shock.

4) Dust from the ambient medium may survive passage through the
   forward bow shock.  For the Trapezium dust arcs, this would require
   dust formation in the stellar wind of $\theta^1$C itself.

Dust is often thought to be destroyed in shocks, giving rise to bright
IR emission lines of [Fe~{\sc ii}] as iron is liberated into the gas
phase (e.g., McKee et al.\ 1984; Hartigan et al.\ 2004; Nisini et al.\
2002; Smith et al.\ 2004a), but the dust may reform in the dense
post-shock cooling zone.  On the other hand, Mouri \& Taniguchi (2000)
have argued that dust may even survive passage through shocks, as
large grains are shattered into smaller grains but are not completely
destroyed.  While options 3 and 4 might contribute some of the dust
seen in HH jets and dust arcs, dust formation or survival in shocks
cannot be the sole explanation for the dust seen in the HH jets in the
Orion nebula.  This is because dust is not only seen at the location
of the bow shock, but is also seen in condensations along the body of
the jet.  This is most clearly illustrated in HH~529 (Fig.\ 7), where
dust is seen from all the main locations along the jet where dense
ionized gas is seen.  HH~514 also shows clear evidence for dust along
the body of the jet (Fig.\ 3).

Option 2 seems to be an unlikely explanation for the source of the
dust as well, but is difficult to rule out.  In the harsh irradiated
environment of the Orion nebula, disks are exposed and no longer
embedded in surrounding cloud cores.  This is especially true for
HH~514 and 513, because they originate from the exposed sources HST2
and 165-235 (proplyds inside the H~{\sc ii} region, with little
surrounding material to entrain), or for the dust arcs around the
proplyds so close to $\theta^1$C.  For these proplyds, dust would need
to be entrained from material in the optically thin proplyd envelope.
However, this process may be important for the embedded sources that
drive HH~202 and HH~529 (Smith et al.\ 2004; Doi et al.\ 2004; O'Dell
\& Doi 2003; Bally et al.\ 2000).

Thus, option 1 above seems to be a likely explanation for the origin
of the dust in HH jets and potentially in the Trapezium dust arcs as
well.  In HH jets, if the dust is entrained directly from material in
the circumstellar disk, one would expect the entrainment to occur at
relatively large radii in the disk -- radii at least as large as the
dust sublimation radius, typically at $R=0.8 (L/L_{\odot})^{1/2}$ \ AU
for $Q_{\rm abs}/Q_{\rm em}$=100 and a sublimation temperature of
$\sim$1000 K.  The requirement that some of the jet material is lifted
from the disk at relatively large radii would provide important
constrains on models of protostellar jet launching and collimation.
The origin of dust in HH jets certainly deserves further
investigation.

In both jets and photoevaporative flows from proplyds, it is likely
that primarily the smallest grains are entrained.  Small grains are
also more likely to survive shocks and the harsh UV radiation field
because of their smaller cross-section, and may be more subject to
stochastic heating.  Measurements of the dust temperature using
multiwavelength data may be able to constrain the grain size
distribution in these HH jets and dust arcs.

\subsection{A New Interpretation of the Ney-Allen Nebula}

We have suggested a new interpretation for the origin of the brightest
arc in the Ney-Allen nebula.  Based on the observed dust color
temperature and its peak near $\theta^1$D, the heating of the Ney
Allen arc appears to be dominated by $\theta^1$D instead of the more
luminous star $\theta^1$C, requiring that $\theta^1$D is actually
behind the other members of the Trapzium, much closer to the
background ionization front and PDR of the Orion nebula.  This
geometry is supported by an absorption feature seen in the spectrum of
$\theta^1$D that is not seen in the other Trapezium stars.  A small
dust-free cavity around $\theta^1$D itself provides additional
evidence that it is embedded in a more dusty environment.

The geometry of the Ney-Allen arc is more puzzling than the question
of its heat source.  One potential explanation may be that the stellar
wind from $\theta^1$D is sweeping up dust from the surrounding PDR.
The proper motion vector of $\theta^1$D makes this suggestion seem
attractive, but is not essential since the wind speed is much greater
than stellar motion.  Although the proper motion vector is small and
highly uncertain, the main point that $\theta^1$D is closer to the PDR
than the other Trapezium stars does not rely on the proper motion
measurement.

\subsection{Future Work}

Figures 8, 9, and 10 reveal a strong spatial anti-correlation between
known proplyds and other IR point sources.  We have argued, based on
the absence of silhouette disks in the ``naked'' stars, that this is
not a simple effect of stronger UV radiation causing enhanced
evaporation in proximity to $\theta^1$C.  Instead, it seems to be
related to age segregation due to subclustering within the ONC,
reflecting the geometry of the cloud cores from which these stars were
born.  This illuminates the vexing question of why proplyds are so
ubiquitous near $\theta^1$C, but are not seen in large numbers
anywhere else in the sky.  More in-depth study at thermal-IR
wavelengths --- especially sensitive multi-filter photometry or
spectroscopy of point sources at 8--50 $\micron$ with ground-based
observatories, {\it Spitzer}, and SOFIA --- will help characterize the
luminosity, dust temperature, dust composition, disk mass, and
extinction as a function of position and specifically as a function of
separation from $\theta^1$C.  It will be important to examine the
emission properties of proplyds and non-proplyds over a larger field
of view, to provide a more complete sample of the inner ONC.

\acknowledgements  \scriptsize

Support for N.S.\ was provided by NASA through grant HF-01166.01A from
the Space Telescope Science Institute, which is operated by the
Association of Universities for Research in Astronomy, Inc., under
NASA contract NAS~5-26555.  Additional support was provided by NSF
grant AST~98-19820 and NASA grants NCC2-1052 and NAG-12279 to the
University of Colorado.



\begin{deluxetable}{llclc}
\tabletypesize{\scriptsize}
\tighten
\tablewidth{0pt}
\tablenum{1}
\tablecaption{11.7 $\micron$ Point Sources in Orion (Excluding BN/KL)}
\tablehead{
 \colhead{R.A.} &\colhead{DEC} &\colhead{F$_{\nu}$(11.7 $\mu$m)} &\colhead{Name/Comment} &\colhead{R05} \\
 \colhead{(J2000)} &\colhead{(J2000)} &\colhead{(Jy)} &\colhead{\ } &\colhead{\ }
}
\startdata

5:35:11.64 &-5:24:21.4  &0.02	&O		&20	\\
5:35:11.95 &-5:22:54.2  &0.08	&X		&25	\\
5:35:12.27 &-5:23:48.3  &0.11	&O; OMC-1 S 8	&27	\\
5:35:12.59 &-5:23:44.3  &0.22	&O; OMC-1 S 7	&31	\\
5:35:12.99 &-5:21:53.2  &0.03	&X		&...	\\
5:35:12.99 &-5:22:15.1  &0.03	&O		&(34)	\\
5:35:13.40 &-5:23:29.5  &0.19	&O; OMC-1 S 10	&40	\\
5:35:13.44 &-5:23:40.4  &0.10	&O; OMC-1 S 9	&41	\\
5:35:13.48 &-5:22:19.5  &(0.03) &X              &...	\\
5:35:13.55 &-5:23:59.9  &0.08	&OMC-1 S C	&42	\\
5:35:13.58 &-5:23:55.5  &0.13	&OMC-1 S B	&43	\\
5:35:13.72 &-5:22:21.9  &(0.1)  &X              &...	\\
5:35:13.72 &-5:22:17.4  &0.12	&X		&...	\\
5:35:13.75 &-5:21:59.6  &9.5	&IRc9 (extended; 5$\farcs$3 diam.\ ap.) &49	\\
5:35:13.75 &-5:22:07.0  &0.35	&138-207	&48	\\
5:35:13.80 &-5:23:40.3  &6.4	&OMC-1 S 1	&46	\\
5:35:13.88 &-5:23:57.4  &0.04	&OMC-1 S 6	&(47)	\\
5:35:14.25 &-5:22:04.5  &0.05	&O		&(55)	\\
5:35:14.27 &-5:23:04.3  &0.13	&X		&58	\\
5:35:14.29 &-5:23:08.4  &0.17	&X		&59	\\
5:35:14.34 &-5:22:54.1  &0.08	&X		&60	\\
5:35:14.39 &-5:23:33.8  &0.28	&OMC-1 S 11	&62	\\
5:35:14.40 &-5:23:51.0  &2.5	&OMC-1 S 2	&61	\\
5:35:14.54 &-5:23:56.2  &0.46	&OMC-1 S 3	&64	\\
5:35:14.67 &-5:22:49.5  &0.06	&X		&(65)	\\
5:35:14.71 &-5:23:23.0  &$<$0.02&147-323	&...	\\
5:35:14.71 &-5:23:14.6  &0.12	&X		&...	\\
5:35:14.86 &-5:23:05.1  &0.04	&O		&66	\\
5:35:14.89 &-5:22:39.2  &0.45	&O		&69	\\
5:35:14.91 &-5:23:29.1  &(0.03) &149-329	&...	\\
5:35:15.17 &-5:22:54.3  &0.59	&O		&71	\\
5:35:15.18 &-5:22:36.8  &0.16	&X		&70	\\
5:35:15.21 &-5:21:55.7  &0.02	&X		&...	\\
5:35:15.30 &-5:22:15.5  &0.46	&O		&73	\\
5:35:15.34 &-5:22:25.2  &0.05	&O		&74	\\
5:35:15.44 &-5:23:45.7  &0.02	&O		&76	\\
5:35:15.46 &-5:22:48.6  &0.04	&O		&...	\\
5:35:15.52 &-5:22:46.5  &0.03	&X		&...	\\
5:35:15.52 &-5:23:37.5  &0.09	&155-338	&79	\\
5:35:15.61 &-5:22:56.5  &1.11	&O		&82	\\
5:35:15.62 &-5:24:03.2  &0.10	&X		&81	\\
5:35:15.72 &-5:23:22.6  &0.04	&157-323	&83	\\
5:35:15.75 &-5:23:10.0  &0.03	&$\theta^1$E	&...	\\
5:35:15.78 &-5:23:26.7  &0.11	&LV 6		&86	\\
5:35:15.80 &-5:23:14.5  &0.05	&$\theta^1$A	&85	\\
5:35:15.81 &-5:22:45.8  &0.04	&O		&84	\\
5:35:15.83 &-5:23:22.5  &0.92	&LV 5		&87	\\
5:35:15.84 &-5:23:25.7  &0.04	&158-326	&(88)	\\
5:35:15.86 &-5:23:02.0  &$<$0.02&O		&...	\\
5:35:15.90 &-5:23:38.1  &$<$0.02&159-338	&...	\\
5:35:15.90 &-5:22:21.0  &0.11	&O		&(91)	\\
5:35:15.95 &-5:23:50.1  &1.70	&HST 3		&...	\\
5:35:15.97 &-5:23:49.7  &(0.03)	&HST 3b		&...	\\
5:35:16.01 &-5:23:53.1  &0.31	&160-353	&94	\\
5:35:16.06 &-5:23:24.4  &0.56	&LV 4		&98	\\
5:35:16.06 &-5:22:54.2  &0.20	&X		&96	\\
5:35:16.06 &-5:23:07.1  &1.29	&$\theta^1$B companion, 161-307 &...	\\
5:35:16.11 &-5:23:06.9  &(0.03)	&$\theta^1$B	&...	\\
5:35:16.23 &-5:22:10.4  &0.21	&O		&102	\\
5:35:16.28 &-5:23:16.3  &0.32	&LV 3		&(105)	\\
5:35:16.32 &-5:23:22.6  &1.63	&163-323; SC3; near $\theta^1$C &106	\\
5:35:16.32 &-5:22:49.1  &0.08	&X		&104	\\
5:35:16.37 &-5:22:12.0  &0.29	&X		&...	\\
5:35:16.39 &-5:24:03.5  &0.30   &O		&(108)	\\
5:35:16.45 &-5:22:35.3  &(0.04)	&HH~513         &...	\\
5:35:16.46 &-5:22:35.1  &(0.09)	&165-235        &...	\\
5:35:16.46 &-5:23:23.0  &0.76	&$\theta^1$C	&111	\\
5:35:16.61 &-5:23:16.2  &0.13	&166-316	&(114)	\\
5:35:16.71 &-5:22:31.2  &0.05	&167-231; silhouette disk &...	\\
5:35:16.74 &-5:23:16.5  &0.96	&LV 2		&116	\\
5:35:16.76 &-5:23:28.0  &0.25	&168-328	&...	\\
5:35:16.84 &-5:23:26.2  &5.59	&LV 1		&118	\\
5:35:16.95 &-5:22:48.5  &$<$0.02&170-249	&(119)	\\
5:35:16.96 &-5:23:00.9  &0.12	&O		&122	\\
5:35:16.98 &-5:23:37.1  &0.39	&HST 2		&121	\\
5:35:17.06 &-5:23:39.8  &0.08	&HST 11		&123	\\
5:35:17.07 &-5:23:34.0  &0.85	&170-334	&124	\\
5:35:17.25 &-5:23:16.6  &0.06	&$\theta^1$D	&...	\\
5:35:17.33 &-5:22:35.7  &0.28	&174-236	&128	\\
5:35:17.35 &-5:22:45.7  &(0.03) &X		&...	\\
5:35:17.39 &-5:24:00.3  &(0.1)  &X		&129	\\
5:35:17.46 &-5:23:21.1  &0.12	&O		&130	\\
5:35:17.49 &-5:21:45.4  &0.22	&O		&...	\\
5:35:17.54 &-5:22:56.7  &0.08	&O		&(132)	\\
5:35:17.56 &-5:21:53.8  &0.07	&X		&(134)	\\
5:35:17.57 &-5:22:56.4  &(0.03)	&O		&...	\\
5:35:17.57 &-5:23:25.0  &(0.02)	&176-325	&131	\\
5:35:17.83 &-5:22:02.8  &0.09	&X		&...	\\
5:35:17.90 &-5:22:44.8  &$<$0.02&O		&140	\\
5:35:18.06 &-5:24:03.1  &0.46	&O		&143	\\
5:35:18.21 &-5:23:35.9  &0.78	&O		&(146)	\\

\enddata \tablecomments{The sources with an ``O'' listed in the right
column are optically visible stars with no extended proplyd structure
in {\it HST} images (O'Dell \& Wong 1996) that nevertheless show
detectable 11.7$\micron$ thermal-IR emission from unresolved dusty
disks.  ``X'' denotes an IR point source with no star or proplyd in
the list of O'Dell \& Wong (1996), and the stars $\theta^1$ABCDE,
members of OMC-1 South, and known proplyds are identified by name.
The last column (R05) gives the designation for sources previously
detected by Robberto et al.\ (2005); sources in parenthesis are
uncertain identifications due to poor agreement in the coordinates.}
\end{deluxetable}


\begin{deluxetable}{lcc}
\tabletypesize{\scriptsize}
\tighten\tablenum{2}\tablewidth{0pt}
\tablecaption{Summary of Visual Sources in IR Survey Region}
\tablehead{
  \colhead{Parameter} &\colhead{Partial Map} &\colhead{Full Map}
}
\startdata

Total number of IR sources detected	&87(+1)	&91(+1)	\\

IR-detected proplyds			&25	&27	\\
Known proplyds				&30	&34	\\
\% proplyds detected in mid-IR		&83\%	&79\%	\\

IR-detected ``naked'' stars		&30	&34	\\
Stars with visual ID			&78	&94	\\
\% visual stars detected in mid-IR	&38\%	&36\%	\\
\% excluding $\theta^1$ABCDE		&34\%	&33\%	\\

IR-detected proplyds \& stars		&55	&61	\\
Total proplyds \& stars			&108	&128	\\
\% of all proplyds \& stars		&51\%	&48\%	\\
\% excluding $\theta^1$ABCDE		&49\%	&46\%	\\

\enddata
\tablecomments{This table compares numbers of sources we detected at
  11.7~$\micron$ with proplyds and stars in our survey region that
  have been identified in visual-wavelength {\it HST} images, listed
  in the catalog of O'Dell \& Wong (1996).  The second column (partial
  map) excludes the region at the left of Figure 1 that is affected by
  sky-chopping artifacts, and so the statitstics are more reliable
  than in the third column, which lists numbers for the full region
  displayed in Figure 1.  Non-proplyd stars (``naked'' stars) are
  those listed as ``objects that appear entirely stellar'' in O'Dell
  \& Wong (1996). This table excludes embedded IR sources that we
  detected at 11.7~$\micron$, which have no visual I.D. The (+1) in
  the first entry referes to the proplyd 141-301, which is not a point
  source.}
\end{deluxetable}

\begin{figure}
\epsscale{0.95}
\plotone{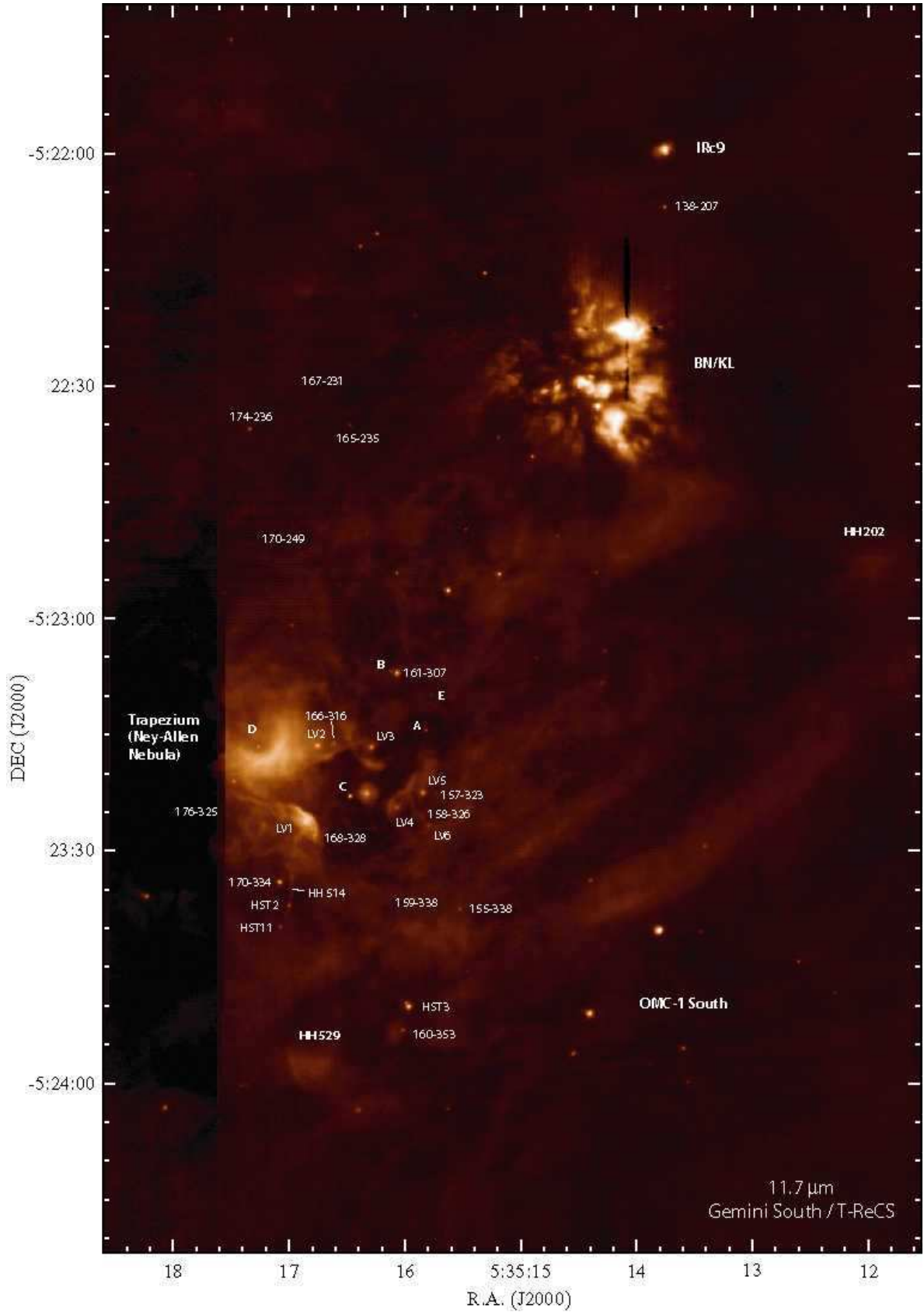}
\caption{11.7 $\micron$ image mosaic of the Orion nebula made with
T-ReCS on Gemini South.  Proplyds seen in {\it HST} images are
identified, as well as a few other major features of the Orion nebula.
The 15$\arcsec$-wide section at the middle of the left side of the
image is an artifact caused by over-subtraction of bright diffuse
emission included in the reference sky beam.}
\end{figure}

\begin{figure}
\epsscale{0.95}
\plotone{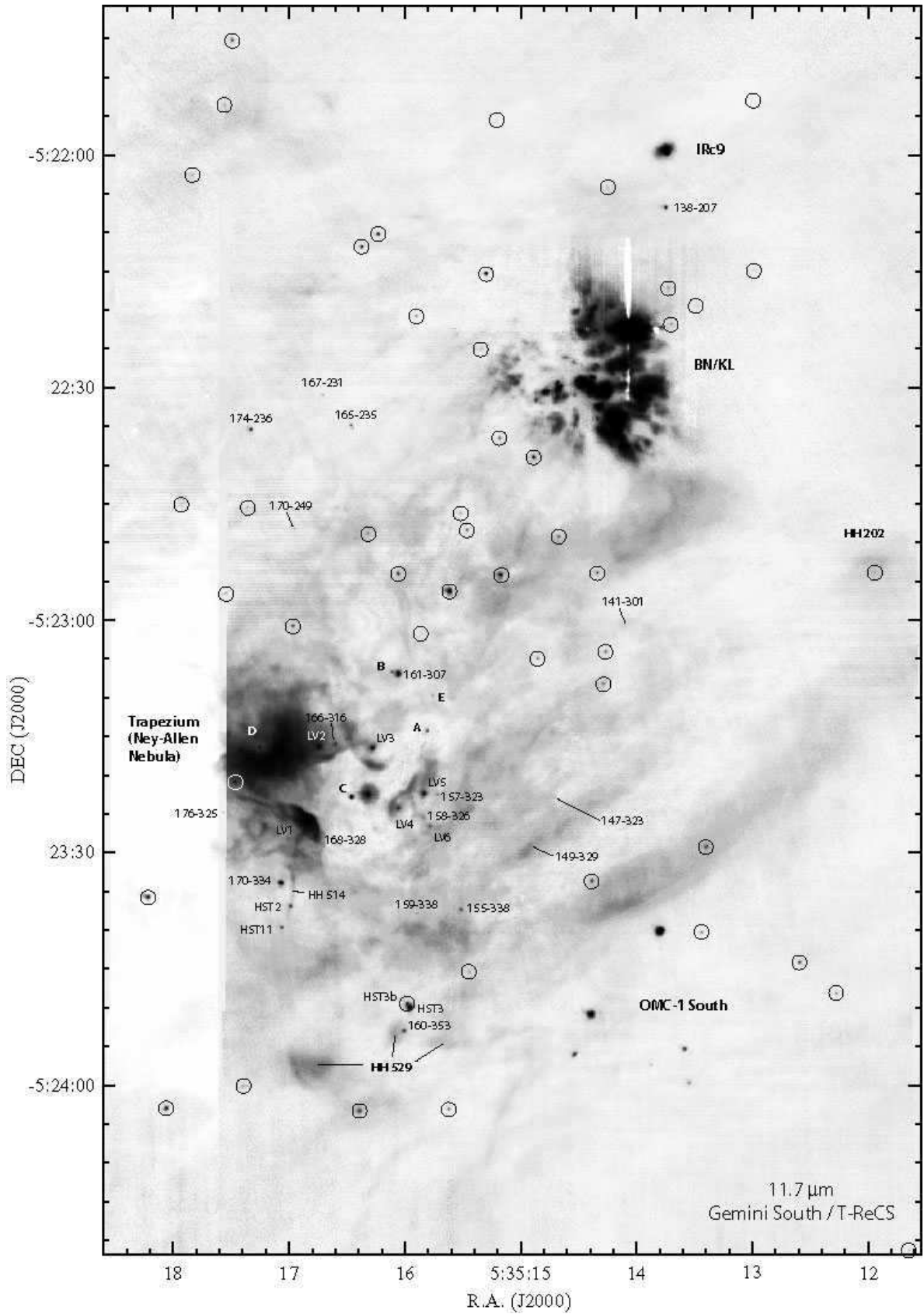}
\caption{Same as Figure 1, but with a deeper grayscale to show fainter
  emission, and with faint and unresolved (non-proplyd) infrared stars
  circled.}
\end{figure}

\begin{figure}
\epsscale{0.95}
\plotone{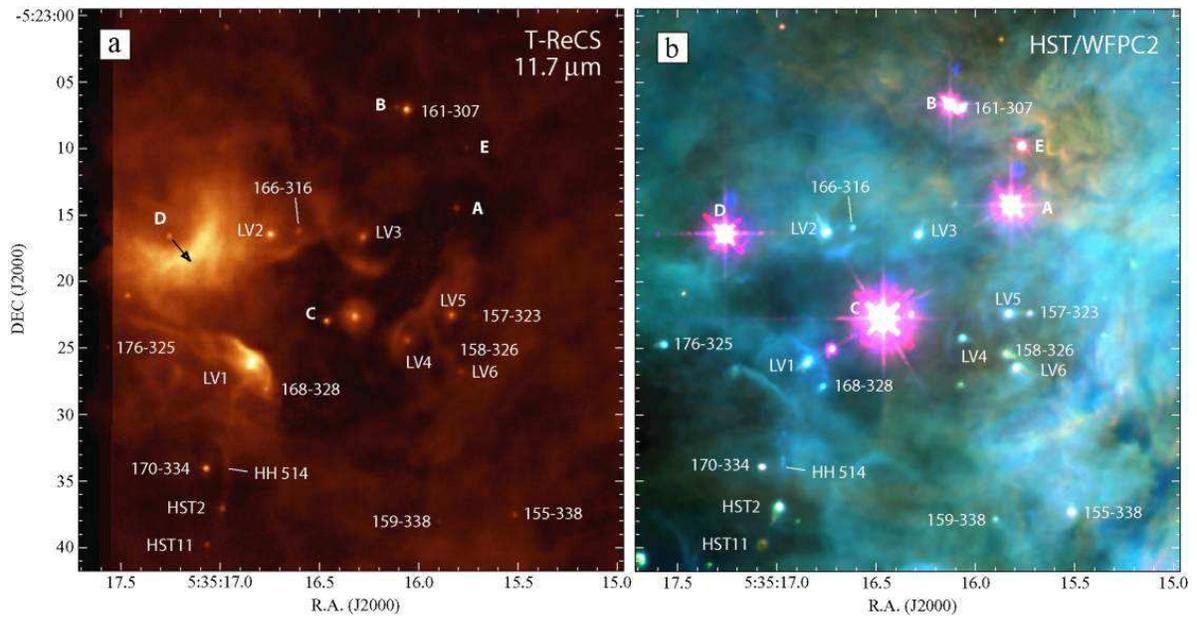}
\caption{A close-up of the Trapezium region (a) at 11.7 $\micron$ with
  T-ReCS, and (b) in H$\alpha$ with {\it HST}/WFPC2.  In the color
  version available in the electronic edition, the 11.7~$\micron$
  image is in false color, while the {\it HST}/WFPC2 image has [O~{\sc
  iii}] in blue, H$\alpha$ in green, and [N~{\sc ii}] in red.}
\end{figure}

\begin{figure}
\epsscale{0.75}
\plotone{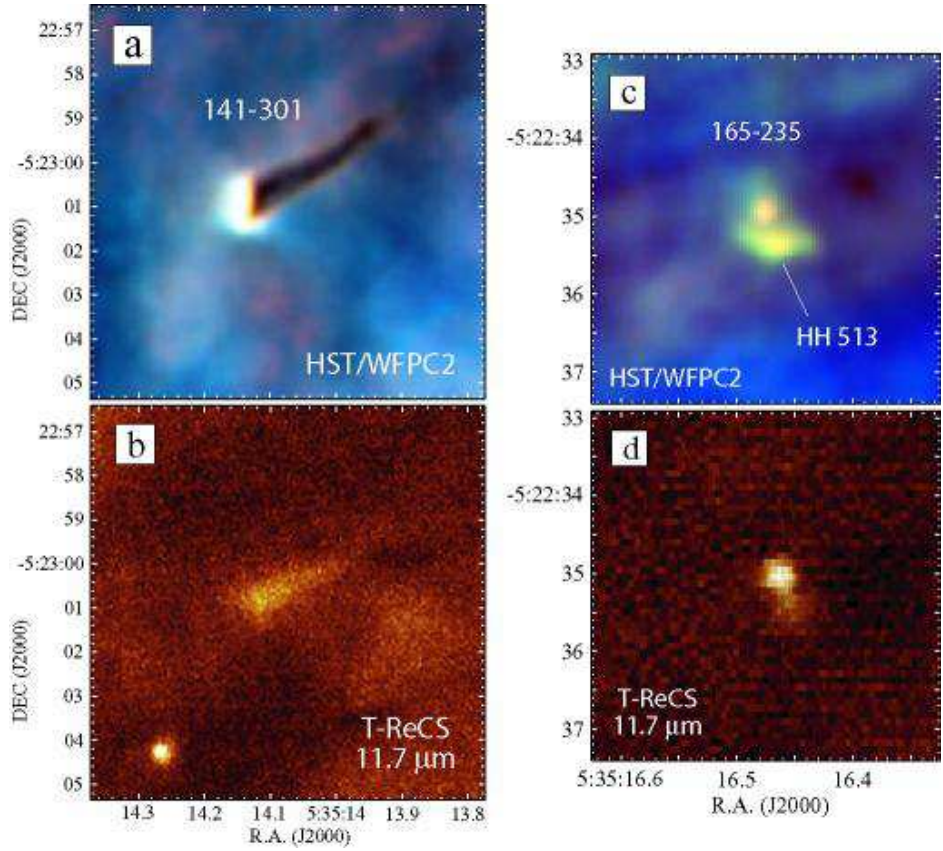}
\caption{Two unusual proplyds more than 30\arcsec\ from $\theta^1$C.
  {\it LEFT}: The unusual proplyd 141-301 (a) seen both as a
  bright-rimmed and silhouette object with {\it HST}/WFPC2 in
  H$\alpha$ and (b) in diffuse emission at 11.7 $\micron$. {\it
  RIGHT}: The proplyd 165-235 and the HH 513 microjet (c) with {\it
  HST}/WFPC2 in H$\alpha$ and (d) at 11.7 $\micron$.  In the color
  figure in the electronic edition, both of the {\it HST} images are
  colored with [O~{\sc iii}] in blue, H$\alpha$ in green, and [N~{\sc
  ii}] in red.}
\end{figure}

\begin{figure}
\epsscale{0.42}
\plotone{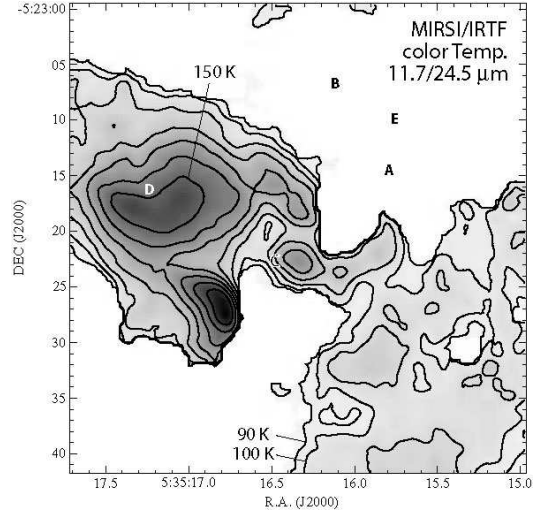}
\caption{An 11.7/24.5 $\micron$ color temperature map of the Trapezium
  obtained with MIRSI at the IRTF (see Kassis et al.\ 2005), showing
  that the local dust temperature peaks in the Ney-Allen arc near
  $\theta^1$D.  Contours of the dust color temperature are drawn at
  increments of 10 K, between 90 and 160 K.  Low surface brightness
  areas in the original images were masked-out before the image ratio
  was taken.  The spatial resolution of this IRTF image is about
  1$\farcs$5, so the temperature structure is not resolved for some
  narrow features seen in our higher-resolution Gemini images.}
\end{figure}

\begin{figure}
\epsscale{0.35}
\plotone{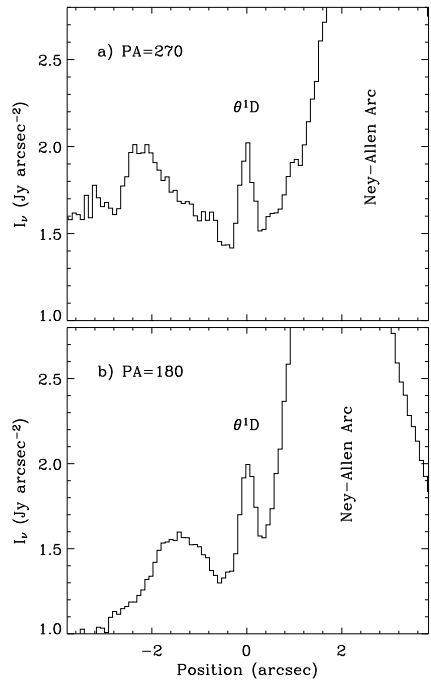}
\caption{A 0$\farcs$4-wide intensity tracing through the position of
  $\theta^1$D in the 11.7~$\micron$ image at two different position
  angles: (a) a horizontal tracing E to W, and (b) a vertical tracing
  N to S.  $\theta^1$D is in the middle, and the strong emission to
  the right in each plot is from the bright extended arc in the
  Ney-Allen nebula.}
\end{figure}

\begin{figure}
\epsscale{0.95}
\plotone{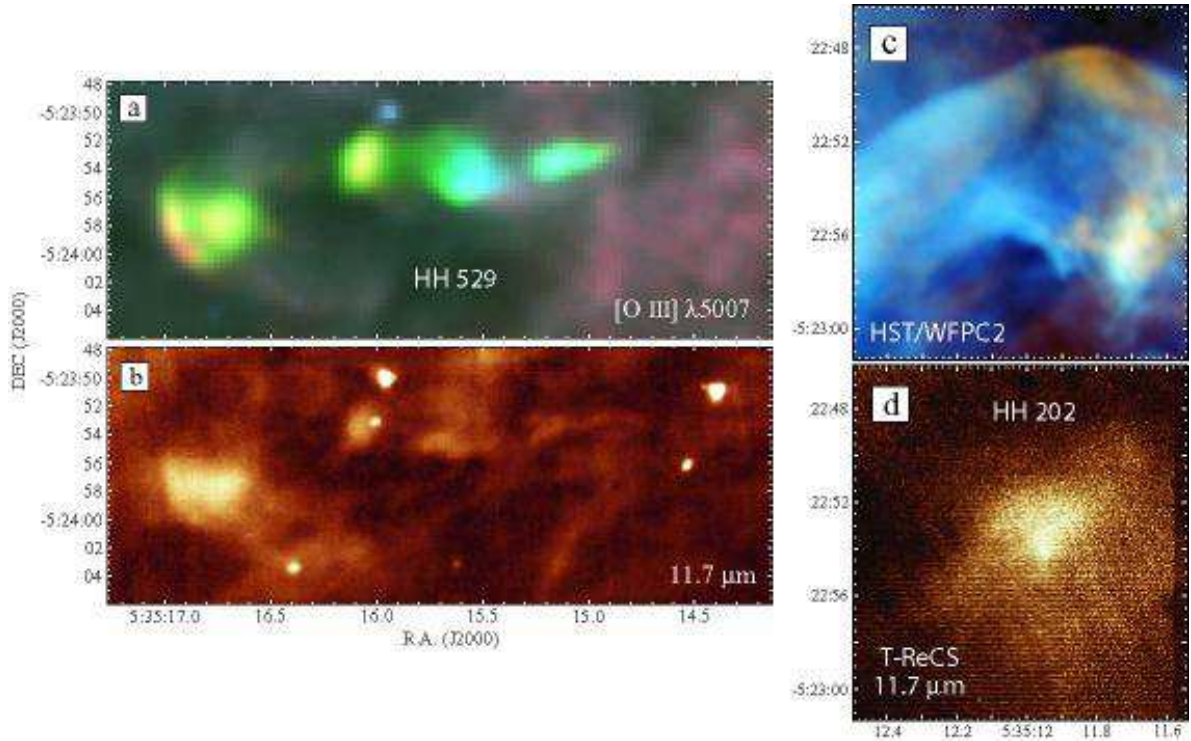}
\caption{{\it LEFT}: A close-up of HH 529 (a) in the [O~{\sc iii}]
  $\lambda$5007 emission line obtained with a ground-based Fabry-Perot
  camera, and (b) at 11.7 $\micron$ with T-ReCS.  In the color version
  available in the electronic edition, the Fabry-Perot image in Panel
  a is color coded by velocity, so that slow blueshifted material is
  red, and progressively faster blueshifted gas is green and then blue
  (see Smith et al.\ 2004a). {\it RIGHT}: The bow shock of HH 202 (c)
  with {\it HST}/WFPC2 in H$\alpha$ and (d) at 11.7 $\micron$.  In the
  color figure in the electronic edition, the {\it HST} image of
  HH~202 is colored with [O~{\sc iii}] in blue, H$\alpha$ in green,
  and [N~{\sc ii}] in red.}
\end{figure}

\begin{figure}
\epsscale{0.85}
\plotone{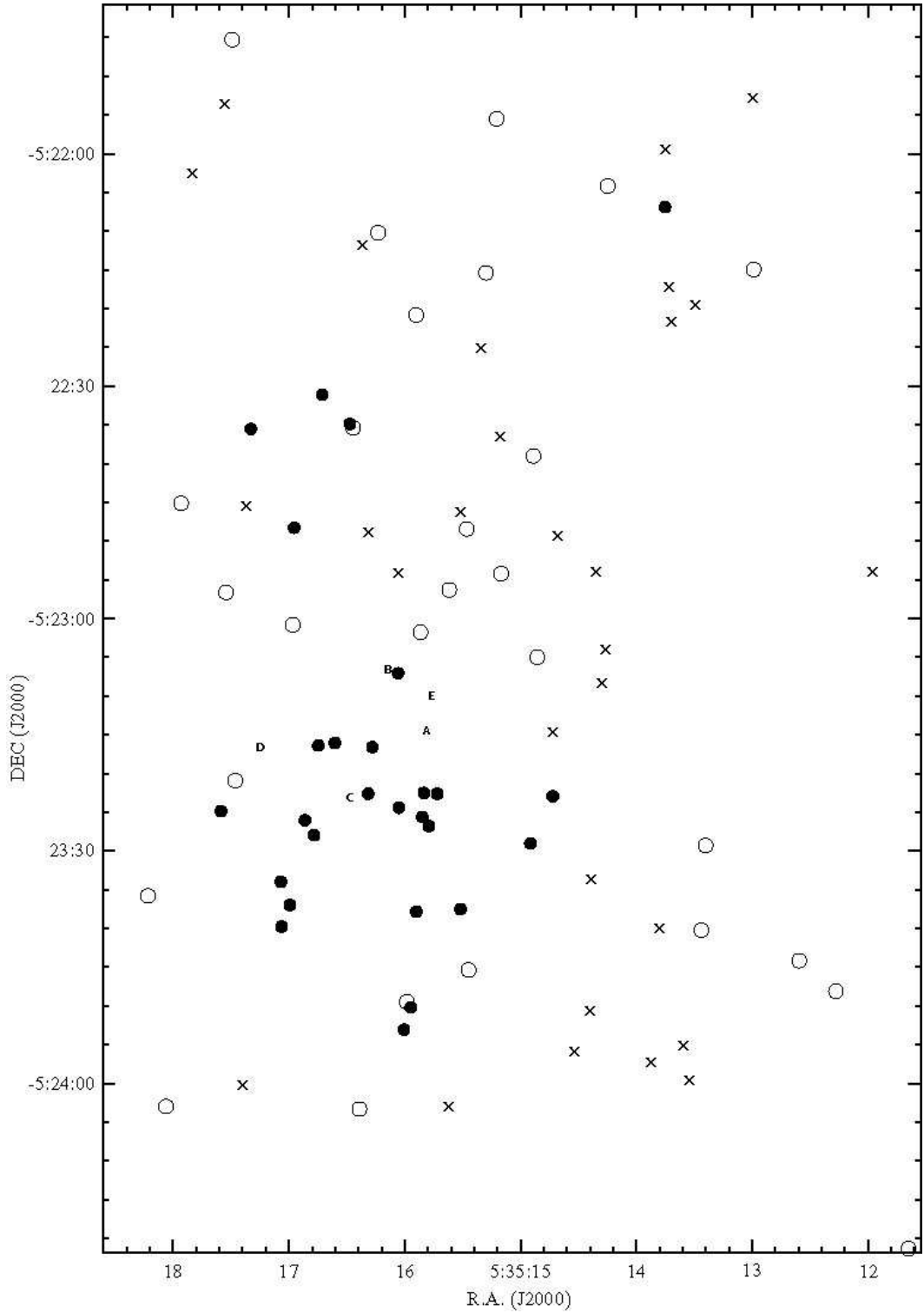}
\caption{Spatial positions of 11.7~$\micron$ point sources that are
  known proplyds (filled dots), naked stars detected at 11.7~$\micron$
  (unfilled circles), and embedded IR sources with no optical ID
  (X's).  The positions of $\theta^1$ABCDE are also shown.}
\end{figure}

\begin{figure}
\epsscale{0.5}
\plotone{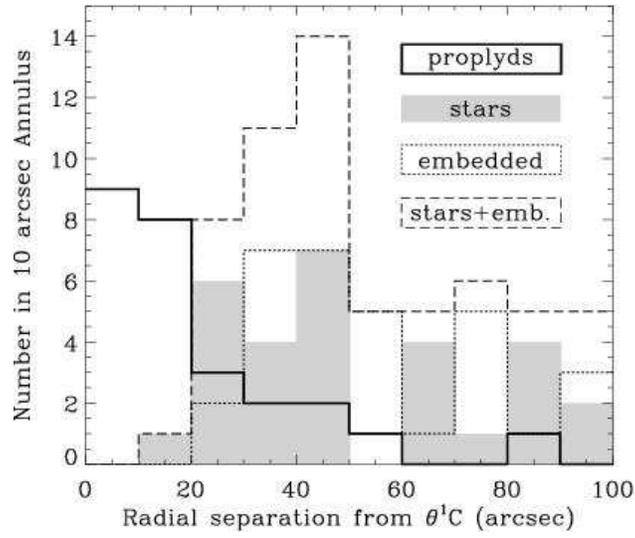}
\caption{Numbers of various sources as a function of projected
  separation from $\theta^1$C, binned in annuli of 10\arcsec.  The
  thick solid histogram shows proplyds detected at 11.7~$\micron$, the
  shaded histogram is for ``naked'' stars, the dotted one is for
  embedded 11.7~$\micron$ point sources, and the dashed histogram is
  the sum of all non-proplyd point sources (``naked'' stars + embedded
  IR sources).  There is a clear anticorrelation in the spatial
  distribution of proplyds and the remaining point sources detected at
  11.7~$\micron$.  This plot excludes $\theta^1$ABCDE.}
\end{figure}

\begin{figure}
\epsscale{0.85}
\plotone{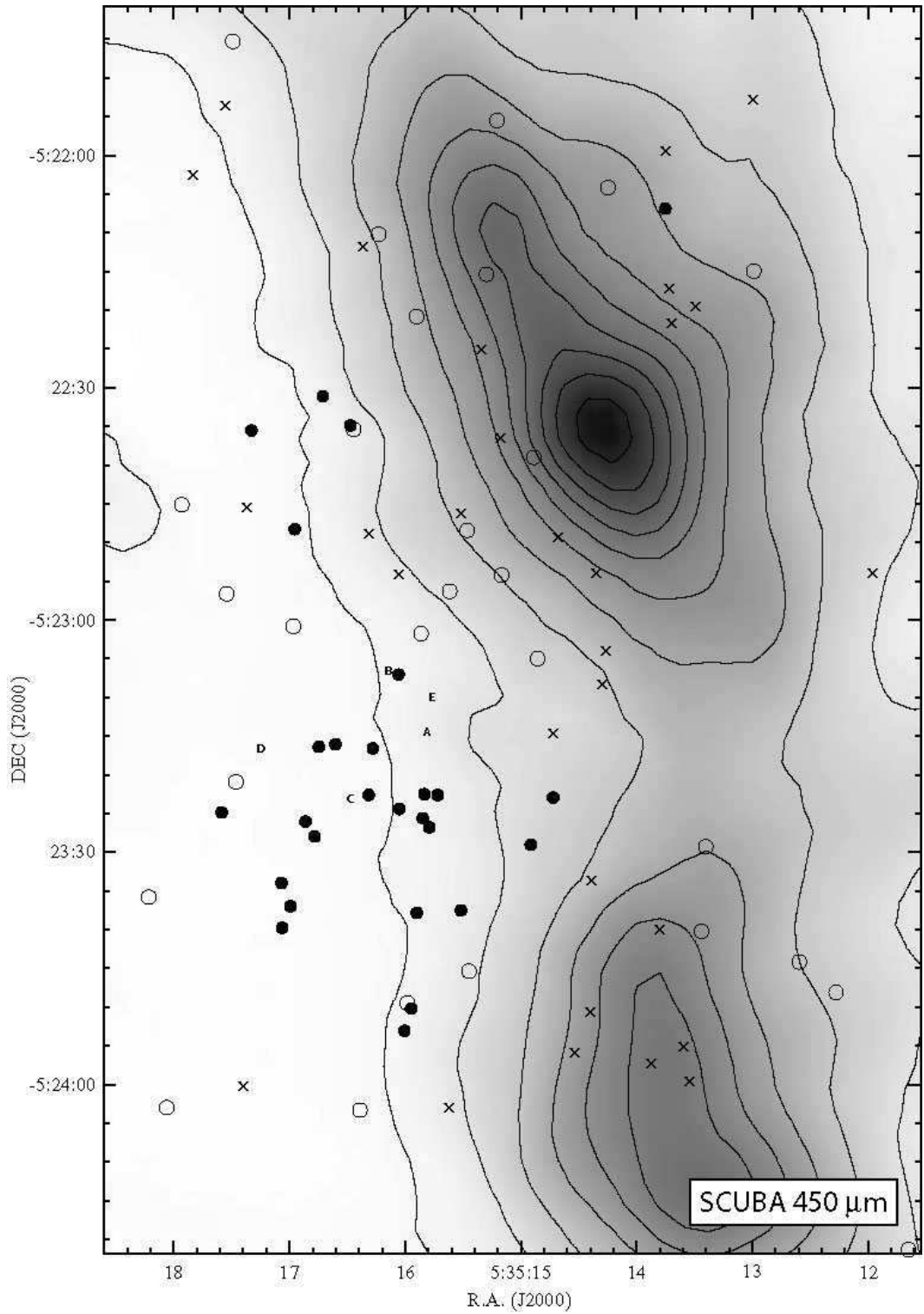}
\caption{Same as Figure 8, but superposed on the 450 $\micron$ SCUBA
  map from Johnstone \& Bally (1999).}
\end{figure}

\begin{figure}
\epsscale{0.45}
\plotone{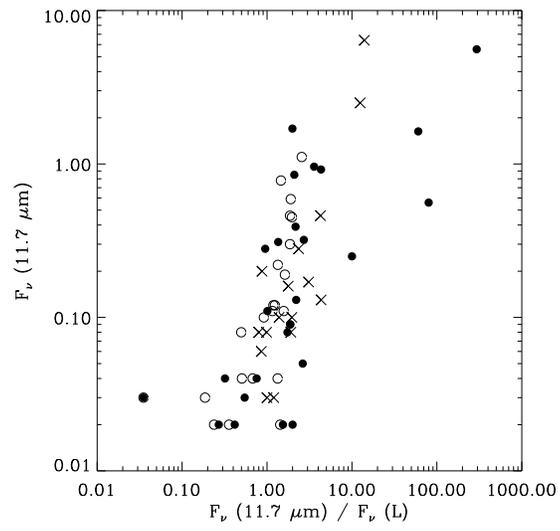}
\caption{An IR color-magnitude diagram for mid-IR detected point
  sources in Table 1.  The L band fluxes are taken from
  Lada et al.\ 2004.  Plotting symbols for proplyds, visible stars
  with unresolved disks, and embedded sources are the same as in
  Figures 8 and 10.}
\end{figure}

\end{document}